\documentclass[nofootinbib,superscriptaddress,amsmath,amssymb]{revtex4-1}

\usepackage{dcolumn}
\usepackage{bm}
\usepackage{amsmath}
\usepackage{amsfonts}
\usepackage{amssymb}
\usepackage[english]{babel}
\usepackage[latin1]{inputenc}
\usepackage{graphicx}
\usepackage{hyperref}
\usepackage{dsfont}
\usepackage{xcolor}
\usepackage{float}

\definecolor{dgreen}{HTML}{00B666}

\newcommand{\rem}[1]{}

\newcommand{\addressZagreb}{Division of Theoretical Physics, Ru\dj er Bo\v{s}kovi\'{c} Institute, Bijeni\v{c}ka c.54, HR-10002 Zagreb, Croatia}
\newcommand{\addressPI}{Perimeter Institute for Theoretical Physics, 31 Caroline Street North, Waterloo, ON, N2L 2Y5 Canada.}

\hypersetup{
    colorlinks=true,         
    linkcolor=blue,          
    citecolor=blue,        
    urlcolor=Violet             
}

\begin{document}
\title{Vector-like deformations of  relativistic quantum phase-space and relativistic kinematics} 
\author{Niccol\'{o}~Loret}
\affiliation{\addressZagreb}
\author{Stjepan~Meljanac} 
\affiliation{\addressZagreb}
\author{Flavio Mercati}
\affiliation{\addressPI}
\author{Danijel~Pikuti\'c}
\affiliation{\addressZagreb}
\begin{abstract}
We study a family of noncommutative spacetimes constructed by one four-vector. The large set of coordinate commutation relations described in this way includes many cases that are widely studied in the literature. The Hopf-algebra symmetries of these noncommutative spacetimes, as well as the structures of star product and twist, are introduced and considered at first order in the deformation, described by four parameters. We also study the deformations to relativistic kinematics implied by this framework, and calculate the most general expression for the momentum dependence of the Lorentz transformations on momenta, which is an effect that is required by consistency. At the end of the paper we analyse the phenomenological consequences of this large family of vector-like deformations on particles propagation in spacetime. This leads to a set of characteristic phenomenological effects.
\end{abstract}

\maketitle

\section{Introduction}

The lattice approach is a well established regulator for Quantum Field Theory (QFT) calculations, which involves approximating spacetime with a discrete lattice of spacing $\ell$. In this way all internal lines of a Feynman diagram are regularized by a momentum cut-off $M \sim 1/\ell$ (in $\hbar = c =1$ units). In Quantum Gravity, on the other hand, the difficulties posed by the non-renormalizability of perturbative General Relativity appear to be a consequence of the continuum assumption, i.e. that the space of functions over which the path integral is defined includes functions which vary appreciably within arbitrarily small regions. It may be tempting then to introduce a minimal length $\ell$ as an actual property of spacetime, in order to regularize Quantum Gravity. Parameter $\ell$ is  generally assumed to be of the order of Planck length $L_p \sim 1.6 \times 10^{-35}$m, and the related energy scale is $E_P\sim 1.2 \times 10^{28}$eV. However this minimal length-scale\footnote{It is more correct to define $\ell$ as a minimal length-scale related to a minimal area, rather than a proper length, since such an identification is possible only in 1+1D.} $\ell$ cannot be simply introduced as a `lattice spacing' for spacetime, as one does in lattice QCD, because this would break Lorentz invariance. One way to keep a notion of minimal length and a sort of momentum cutoff in a Lorentz-invariant way is to 
implement $\ell$ as a \emph{quantum} property of spacetime. For example, the angular momentum in nonrelativistic quantum mechanics can have a nonzero expectation value without breaking rotational symmetry (in all $m=0$, $l \neq 0$ states)~\cite{GAC-OnTheFate}. This is realized by introducing the deformation scale (in the case of angular momentum, $\hbar$) into the \emph{commutation relations} describing the system. In this paper we assume that the Planck scale is similarly introduced into the quantum structure of spacetime through some nontrivial commutation relation between coordinates describing generalized $\kappa$-deformed Minkowski space \cite{nontrivialcomm}, such as
\begin{equation}
[\hat{x}^\mu , \hat{x}^\nu]= i(a^\mu \hat{x}^\nu - a^\nu \hat{x}^\mu), \label{CommRel1}
\end{equation} 
in which $a^\mu$ is a four vector proportional to $M^{-1}$, where $M$ is a deformation scale that, in units $c=\hslash=1$, has the dimensions of an energy. Describing the symmetries  of such noncommutative spacetimes requires the mathematical framework of Hopf algebras \cite{HopfGeneral1,HopfGeneral2}, which generalize the Lie algebra structure of the special-relativistic Poincar\'{e} group. From the physical point of view, this approach leads to a generalization of Special Relativity realized within the framework of Deformed Special Relativity \cite{DSR}, which inspired a large number of phenomenological investigations seeking experimental evidence of Planck-scale effects \cite{phenomenology,neutrini1,neutrini2}.

In this paper we propose a general study of a large class of noncommutative spacetimes, their relativistic properties and some preliminary investigations on the physical effects that can be expected by the various configurations of vector $a^\mu$, as defined in Refs.~\cite{MeljkMink,EPJC,HermitianRealizations}. We will in fact define the possible classes of deformations as {\it timelike}, {\it lightlike} or {\it spacelike}, according to the value of the norm of $a^\mu$, respectively $a_\nu a^\nu=\{a^2,0,-a^2\}$, $a \in \mathbb{R}$ (in this paper the convention for the Minkowski metric is $\eta=\text{diag}(1,-1,-1,-1)$).

In Section~\ref{sec:vldef} we will introduce the most general vector-like realization of a deformed Heisenberg algebra that can be written using only one vector $a^\mu$ (at all orders in $M^{-1}$). For this purpose, it is useful to split $a^\mu$ into a dimensionless normal vector $u^\mu$ and the scale factor $M^{-1}$, so that $a^\mu \sim u^\mu/M$. Such a general framework contains $\kappa$-Minkowski, one of the most studied noncommutative spacetimes~\cite{kMinkRelation}, as special timelike case at first order in $M^{-1}$. Also Snyder \cite{Snyder,SnyderSFT} commutation relations can be obtained by a particular choice of $\hat{x}$ configuration (of course taking into account corrections of order $M^{-2}$), but not the Moyal one, because its characteristic antisymmetric tensor $\theta^{\mu\nu}$ cannot be reproduced introducing just one vector, and should be then investigated using a different formalism. In Section~\ref{sec:realiz} we will define our deformed Heisenberg algebra at first order in $M^{-1}$, and define the Hopf-algebra structures (left hand side action, star product and coproduct) that are needed to describe its symmetries. Section~\ref{sec:GLT} is dedicated to a more physical exploration of the properties of the noncommutative spacetimes we introduced. We will write a general parametrization for the boost generator and investigate on the relativistic properties of boost transformations. In Section~\ref{sec:Perturb}, we will find some perturbative results up to first order in $1/M$, obtaining some constraints from relativistic principle, useful to discuss some special cases of particular interest in literature \cite{EPJC}. Finally, in Section~\ref{sec:Phenom} , we will explore a large set of phenomenological effects  arising from different noncommutative spacetime models. We will observe that the large parametrization that we are going to introduce in Sec.~\ref{sec:vldef} will allow us to shed more light on the relations between the shape of coordinate noncommutativity and Planck-scale effects on particle propagation in spacetime.

\section{Vector-like deformations}\label{sec:vldef}
The undeformed Heisenberg algebra $\mathcal H(x,p)$ is given by:
\begin{equation}\begin{split}
[x_\mu, x_\nu]=0&\,,\quad [p_\mu, p_\nu]=0\,, \\
[x_\mu, p_\nu]&=i\eta_{\mu\nu}\,.
\end{split}\end{equation}
The undeformed coordinates $x_\mu$ generate an enveloping algebra $\mathcal A(x)$, which is a subalgebra of $\mathcal H(x,p)$, i.e. $\mathcal A(x) \subset \mathcal H(x,p)$. The momenta $p_\mu$ generate the algebra $\mathcal T(p)$, which is also a subalgebra of $\mathcal H(x,p)$, i.e. $\mathcal T(p) \subset \mathcal H(x,p)$. 

The most general realization of noncommutative coordinates $\hat x_\mu$, linear in the commutative coordinates $x^\beta$,\footnote{This assumption ensures the noncommutative spacetime we want to describe to be geometrically flat. By keeping coordinates at most linear in every expression we are mantaining the vector space structure that characterizes the Cartesian coordinates of Minkowski spacetime.} that can be constructed with a vector $u_\mu \in \mathcal M_{1,n-1}$, such that $u^2\in\{1,0,-1\}$, and a dimensionful deformation parameter $M^{-1}$, where $M$ is of order of the Planck mass, is given by:
\begin{equation}\begin{split}
\hat x_\mu &= x_\mu f_1 +\frac{f_2}{M}
u_\mu(x\cdot p) + \frac{f_3}{M}u_\mu(u\cdot x)(u\cdot p) + \frac{f_4}{M}(u\cdot x)p_\mu  +\frac{f_5}{M^2}(x\cdot p)p_\mu +\frac{f_6}{M}u_\mu + \frac{f_7}{M^2}p_\mu =\label{gendef}\\
&=x^\alpha \varphi_{\alpha\mu}\left(\frac pM \right) + \frac1M \chi_\mu \left(\frac pM \right)\,,
\end{split}\end{equation}
where $f_{1,...,7}$ are functions of $A\equiv u\cdot p/M$ and $B\equiv p^2/M^2$. 

The deformed Heisenberg algebra $\hat{\mathcal H}(\hat x,p) \sim \mathcal H(x,p)$ is given by (see also \cite{NCspaces, Mignemi}):
\begin{align}\begin{split}\label{DHAxx}
[\hat x_\mu , \hat x_\nu]&=i(u_\mu \hat x_\nu - u_\nu \hat x_\mu)\frac{F_1}{M} + i(\hat x_\mu p_\nu - \hat x_\nu p_\mu)\frac{F_2}{M} +i(u\cdot\hat x)(u_\mu p_\nu - u_\nu p_\mu)\frac{F_3}{M^2} +\\
&+i(\hat x \cdot  p)(u_\mu p_\nu - u_\nu p_\mu)\frac{F_4}{M^3} +i(u_\mu p_\nu - u_\nu p_\mu)\frac{F_5}{M^3}\,,
\end{split}\\
[\hat x_\mu, p_\nu]&=i\eta_{\mu\nu}f_1+\frac iMu_\mu p_\nu f_2 + \frac iM u_\mu u_\nu (u\cdot p) f_3 +\frac iM u_\nu p_\mu f_4 + \frac i{M^2} p_\mu p_\nu f_5\,, \\
[p_\mu,p_\nu]&=0\,,
\end{align}
where $F_{1,..,5}$ are also functions of $A$ and $B$ which can be expressed in terms of the functions $f_{1,...,7}$ and their derivatives. The calculation of these functions is straightforward but tedious. 
 If we restrict the discussion to the deformations of Minkowski space which form a closed algebra, then from formula \eqref{DHAxx} it follows that only the case of the generalized kappa-deformed Minkowski space~(\ref{CommRel1}) would remain. However, it is important to point out that this construction unifies commutative spaces to various types of non-commutative spaces, including $\kappa$-Minkowski space~\cite{kMinkRelation}, Snyder space~\cite{Snyder}, generalized Snyder spaces~\cite{Mignemi} and $\kappa$-Snyder spaces~\cite{kappaSnyder}. The Moyal space ($\theta$-deformation) and quadratic deformations are not included in this construction.
The coproduct of $p_\mu$, the twist $\mathcal F$ and the star product can be calculated perturbatively, as we will show in the next two sections.

\section{Realizations at first order in $1/M$}\label{sec:realiz}

The vector-like realizations at first order in $1/M$ with $\chi_\mu(p)=0$ are given by:
\begin{equation}\label{linreal}
\hat x_\mu = x_\mu + \frac1M\big(c_1 x_\mu(u\cdot p) + c_2 u_\mu(x\cdot p)+ c_3 u_\mu (u\cdot x)(u\cdot p) + c_4 (u\cdot x)p_\mu \big)  = x_\mu + K_{\beta\mu\alpha}x^\alpha p^\beta \, ,
\end{equation}
where $c_{1,2,3,4}\in\mathbb R$.The commutator of these deformed coordinates is:
\begin{equation} \label{pertxx}
[\hat x_\mu, \hat x_\nu] = i \frac{c_1-c_2}M(u_\mu \hat x_\nu - u_\nu \hat x_\mu) + \mathcal O(1/M^2).
\end{equation}

\subsection{Action $\triangleright$, star product and twist}

The left-action $\triangleright$ of $\mathcal H$ on $\mathcal A$ is a map $\triangleright: \mathcal H(x,p)\otimes \mathcal A(x) \rightarrow \mathcal A(x)$ with the following properties:
\begin{equation}\begin{split}
x_\mu \triangleright f(x) &= x_\mu f(x) \, ,\\
p_\mu \triangleright f(x) &=[p_\mu, f(x)]=-i\frac{\partial f(x)}{\partial x^\mu}\, .
\end{split}\end{equation}
Analogously to $\mathcal A(x)$, the deformed coordinates $\hat x_\mu$ generate an enveloping algebra $\hat{\mathcal A}(\hat x)$ if the commutation relations, Eqs.~\eqref{DHAxx} and~\eqref{pertxx}, are closed in $\hat x_\mu$. For every function $f(x) \in \mathcal A(x)$, which can be Fourier expanded, there exists $\hat f(\hat x) \in \hat{\mathcal A}(\hat x)$ such that~\cite{NCspaces, Mignemi}:
\begin{equation}
\hat f(\hat x) \triangleright 1=f(x).
\end{equation}

The star product $\star:\mathcal A(x)\otimes \mathcal A(x)\rightarrow \mathcal A(x)$ is defined by:
\begin{equation}
f(x)\star g(x) = \left(\hat f(\hat x) \hat g(\hat x)\right)\triangleright 1 = \hat f(\hat x)\triangleright g(x).
\end{equation}
For Lie-algebraic deformations, the star product is associative
\begin{equation}
\left(f(x)\star g(x)\right)\star h(x) = f(x) \star \left(g(x)\star h(x)\right).
\end{equation}

The star product is related to the twist operator $\mathcal F^{-1}\in \mathcal H(x,p) \otimes \mathcal H(x,p)$ constructed in~\cite{twists} in the following way:
\begin{equation}
f(x) \star g(x) = m \left[ \mathcal F^{-1} (\triangleright \otimes \triangleright)(f(x) \otimes g(x)) \right]\, ,
\label{startwist}
\end{equation}
where $f(x),g(x) \in \mathcal A(x)$ and $m:\mathcal H(x,p)\otimes\mathcal H(x,p) \to \mathcal H(x,p)$ is the multiplication map of the algebra $\mathcal H(x,p)$. Furthermore \cite{NCspaces},
\begin{equation}\label{hatfFf}
\hat f(\hat x(x,p)) = m \left[ \mathcal F^{-1} (\triangleright \otimes 1)(f(x) \otimes 1) \right], \quad f(x) \in \mathcal A(x) \,,
\end{equation}
where $\hat f(\hat x) \in \hat{\mathcal A}(\hat x)$ is expressed in terms of $x, p \in \mathcal H(x,p)$. As a special case of equation \eqref{hatfFf}, the deformed coordinates $\hat x_\mu$ can be obtained from the twist using the following relation:
\begin{equation}
\hat x_\mu = m\left[\mathcal F^{-1}(\triangleright\otimes1)(x_\mu\otimes1) \right].
\end{equation}

The twist $\mathcal F^{-1}$ related to linear realization of $\hat x_\mu=x_\mu + K_{\beta\mu\alpha}x^\alpha p^\beta$ is given by \cite{EPJC}:
\begin{equation}
\mathcal F^{-1} = e^{ip^W_\alpha\otimes(\hat x^\alpha - x^\alpha)} = 1\otimes1 + p_\alpha \otimes (\hat x^\alpha - x^\alpha) + \mathcal O(1/M^2)\,,
\end{equation}
where $p^W_\mu$ is a function of $p_\mu$ and $u_\mu$ satisfying:
\begin{equation}
(p^W_\mu-k_\mu)e^{ik\cdot\hat x}\triangleright 1 = 0
\end{equation}
where superscript $W$ denotes Weyl ordering. Note that $\hat x_\mu - x_\mu = K_{\beta\mu\alpha}x^\alpha p^\beta + \mathcal O(1/M^2)$.

The coproduct $\Delta p_\mu$ is given by
\begin{equation} \label{Deltapmu}
\Delta p_\mu = \mathcal F \Delta_0 p_\mu \mathcal F^{-1} = p_\mu \otimes 1 + \left(e^{\mathcal K} \right)_{\mu\alpha}\otimes p^\alpha=\Delta_0 p_\mu - K_{\mu\alpha\beta}p^\alpha\otimes p^\beta + \mathcal O(1/M^2)\,,
\end{equation}
where $\Delta_0 p_\mu = p_\mu \otimes 1 + 1 \otimes p_\mu$ and $\mathcal K = -K_{\mu\alpha\nu}(p^W)^\alpha$. 

The star product between two plane waves gives the deformed addition of momenta \cite{EPJC, KovMel}:
\begin{equation}
e^{ik\cdot x}\star e^{iq\cdot x}=e^{i\mathcal D(k,q)\cdot x} \,,
\end{equation}
where $k_\mu, q_\mu \in \mathcal M_{1,n-1}$ and the function $\mathcal D_\mu(k,q)= (k\oplus q)_\mu$ satisfies
\begin{equation}
\Delta p_\mu = \mathcal D_\mu(p\otimes 1, 1\otimes p)\label{comp&copr},
\end{equation}
in agreement with \eqref{Deltapmu}. The function $\mathcal D_\mu(k,q)$ describes the deformed addition of momenta $(k\oplus q)_\mu=\mathcal D_\mu(k,q)$.

\section{General Lorentz transformations}\label{sec:GLT}
The mathematical construction introduced in Sections~\ref{sec:vldef} and~\ref{sec:realiz} has profound implications for the relativistic framework of physical models.
The deformed commutation relations we introduced above are not necessarily Lorentz-invariant in the ordinary sense. The introduction of the energy scale $M$ into the structure of space time/phase space is incompatible with standard Lorentz relativistic structure, but, as is well-known after a few decades of studies of noncommutative spacetimes, this does not imply that the relativistic equivalence between inertial frames is lost. In many cases, a `deformed'  action of the Lorentz group allows to restore relativistic invariance~\cite{amelino2012fate,FlaJosePepe}.
A common assumption in this framework is that the Lorentz group itself (\emph{id est} the commutation relations between the $M_{\mu\nu}$ generators) is not deformed (this is justified by the fact that a dimensionful parameter like $M^{-1}$ cannot enter the algebraic structure of the Lorentz group). What is deformed is the action of the group on noncommutative coordinates $\hat x_\mu$ and momenta $p_\mu$, and on composed momenta (see below). Typically, there are several possible realizations of the action of the Lorentz group on our deformed phase space~\cite{KMPS}, and we want to `parametrize our ignorance' by writing the most generic one and then constraining its free parameters. To this end, we employ the trick of introducing a nonlinear realization of Heisenberg's algebra in which the new set of coordinates $(X_\mu$, $P_\nu)$ generates an undeformed Heisenberg algebra $\mathcal H(X,P)\sim\mathcal H(x,p)$:
\begin{equation}\begin{split}
[X_\mu, X_\nu]=0&, \quad [P_\mu, P_\nu]=0,\label{relnewcoord} \\
[X_\mu, P_\nu]&=i\eta_{\mu\nu}\,,
\end{split}
\end{equation}
but the generators $X_\mu$, $P_\mu$ are related to $x_\mu$, $p_\mu$ by a similarity transformations \cite{KMPS}:
\begin{align}
P_\mu &= \Sigma_\mu(p), \label{Pmu}\\
X_\mu &= x^\alpha \psi_{\alpha\mu}(p),\label{Xmu}
\end{align}
satisfying
\begin{equation}
\frac{\partial\Sigma_\mu(p)}{\partial p_\alpha}\psi_{\alpha\nu}=\eta_{\mu\nu}.\label{dSigma}
\end{equation}
 The coordinate commutativity is automatically satisfied by relations (\ref{Pmu}-\ref{dSigma}). To see this, consider the following. Starting with equations (\ref{Pmu}) and (\ref{Xmu}), let us assume that $X_\mu$ and $X_\nu$ do not commute. Then,
\begin{equation}
 \left[ X_\mu, X_\nu \right] = [x^\alpha \psi_{\alpha\mu}(p) , x^\beta \psi_{\beta\nu}(p)] =i \, x^\alpha  \left( \frac{\partial \psi_{\alpha\nu}}{\partial p_\beta} \psi_{\beta\mu}  - \frac{\partial \psi_{\alpha\mu}}{\partial p_\beta} \psi_{\beta\nu} \right) = X^\lambda C_{\mu\nu\lambda}(p) \,.
\end{equation}
Since we assume that $[X_\mu, P_\nu]=i\eta_{\mu\nu}$ (eq. (\ref{relnewcoord})), then, from Jacobi relations with $X,X,P$, it follows $[[X_\mu, X_\nu], P_\lambda]=0$ and consequently, $C_{\mu\nu\lambda}(p)=0$, which implies that $[X_\mu, X_\nu]=0$.
Note that Eq.~(\ref{dSigma}) implies that the constraint $C_{\mu\nu\lambda}(p)=0$ is automatically satisfied.
The trick we employed allows to introduce a Lorentz generator with the properties that we want at no cost. Because the new basis satisfies undeformed commutation relations, we can introduce a generator of infinitesimal Lorentz transformations as
\begin{equation}
M_{\mu\nu} = X_\mu P_\nu  - X_\nu P_\mu,
\end{equation}
so that the commutation relations between $M_{\mu\nu}$ and itself will close an $so(3,1)$ algebra by construction, and their action  on $X_\mu$ and $P_\mu$ will be undeformed. Therefore the boost representation in terms of the coordinates $(x,p)$ will be:
\begin{equation}
M_{\mu\nu}=x^\alpha(\psi_{\alpha\mu}(p)\Sigma_\nu-\psi_{\alpha\nu}(p)\Sigma_\mu),
\end{equation}
and the commutators of $M_{\mu\nu}$ with $\hat x_\lambda$ and $p_\lambda$ will be of the following form:
\begin{align}
[M_{\mu\nu},\hat x_\lambda]&=\hat x^\alpha \Gamma_{\mu\nu\lambda\alpha}(p),\\
[M_{\mu\nu}, p_\lambda]&=p^\alpha H_{\mu\nu\lambda\alpha}(p),
\end{align}
where $\Gamma_{\mu\nu\lambda\alpha}$ and $H_{\mu\nu\lambda\alpha}$ can be expressed in terms of $\varphi_{\mu\nu}$ and $\Sigma_\mu$.

From the construction of $M_{\mu\nu}$ and the Heisenberg algebra $\mathcal H(X,P)$, it follows that the generators $X_\mu$, $P_\mu$ transform like vectors under $M_{\mu\nu}$:
\begin{align}
[M_{\mu\nu},X_\lambda]&=i(X_\mu\eta_{\nu\lambda} - X_\nu\eta_{\mu\lambda}),\\
[M_{\mu\nu},P_\lambda]&=i(P_\mu\eta_{\nu\lambda} - P_\nu\eta_{\mu\lambda}).
\end{align}

The coproduct $\Delta M_{\mu\nu}$ is $\Delta M_{\mu\nu}=\mathcal F \Delta_0M_{\mu\nu} \mathcal F^{-1}$,
where it is important to note that $\Delta_0 M_{\mu\nu}$ is not primitive, i.e. $\Delta_0 M_{\mu\nu} \ne M_{\mu\nu}\otimes 1 + 1 \otimes M_{\mu\nu}$. Instead, $\Delta_0M_{\mu\nu}(x^\alpha p_\beta, p_\gamma)=M_{\mu\nu}(\Delta_0x^\alpha p_\beta, \Delta_0p_\gamma)$, where $\Delta_0x^\alpha p_\beta$ and $\Delta_0p_\gamma$ are primitive, i.e. $\Delta_0 x^\alpha p_\beta=x^\alpha p_\beta\otimes1+1\otimes x^\alpha p_\beta$ and $\Delta_0 p_\gamma=p_\gamma\otimes1 + 1\otimes p_\gamma$.

The Lorentz transformation of $P_\mu$  with respect to the generator $M_{0i}$ is simply 
\begin{equation}\label{UndeformedBoost}\begin{split}
P'_0&=P_0\cosh\xi  + P_i\sinh\xi, \\
P'_i&=P_i\cosh\xi  + P_0\sinh\xi,
\end{split}\end{equation}
and $P'_j=P_j$ for $j\ne i$.

In order to obtain the form of the boost transformations on $p_\mu$ we express $p_\mu=\Sigma^{-1}_\mu(P)$ as a function of $P_\mu$. Consider an infinitesimal Lorentz transformation (or rotation) of a particle momentum with rapidity parameters $\Xi_{\mu\nu}$ (these are just infinitesimal antisymmetric matrices, usually one has $\Xi_{0i}=\xi_{(i)}$, $\Xi_{ij}|_{i\neq j}=\theta_{(i,j)}$, where the $\xi$s are  parameters with the dimensions of velocity and the $\theta$s are angles). We can calculate its explicit form, at first order in $\Xi_{\mu\nu}$, by using the fact that the generators $P_\mu$ transform classically:
\begin{equation}
P'_\mu = P_\mu + \Xi^{\alpha\beta}[M_{\alpha\beta}, P_\mu] = P_\mu + \Xi_\mu^{\;\beta} P_\beta ,
\end{equation}
then we know what the action on $p_\mu$ is:
\begin{equation}
p'_\rho = \Sigma^{-1}_\rho \left[\Sigma_\mu(p) +\Xi_\mu^{\;\sigma} \Sigma_\sigma (p)\right] ,
\end{equation}
This defines a Lorentz boost on $p$ as a map $\Lambda(\xi,p)$. Expanding this function at first order in $\Xi_{\mu\nu}$:
\begin{equation}\label{TransformationOfPsmall}
p'_\gamma = p_\gamma +\Xi_\rho^{\;\beta} \Sigma_\beta(p)\partial_\rho \Sigma^{-1}_\gamma(p)=\Lambda(\Xi,p).
\end{equation}
Once the form of the boosts is obtained, one can find the relativistic relations between different observers in spacetime. In the low rapidity limit $\xi\ll 1$ the Galileo transformations characterize the relations between energies of two boosted observers. Furthermore Special Relativity also defines the time dilatation and length contraction effects seen by a stationary observer while looking at phenomena characterized by high velocities (i.e. $\xi\gg 1$). However in any case (and in any limit) even if two boosted observers measure different energies, momenta, time-intervals and lengths, the Relativity Principle imposes that if some physical phenomenon, e.g. the decay of a particle, takes place in some reference frame, spacetime transformations should always allow all the observers to agree on the fact that such a decay happened. One consequence of this principle is that  interaction vertices should be conserved by boost transformations. For instance, in the decay of a particle with momentum $p$ in two particles with momenta $k$ and $q$, we expect to have $p'=k' \oplus q'$ , where $p$, $q$ and $k$ are transformed according to~(\ref{TransformationOfPsmall}). However, taking into account that our deformed addition of momenta $(k\oplus q)_\mu=\mathcal D_\mu(k,q)$ can be nonlinear, 
generally 
\begin{equation}
\mathcal D(k,q)' \ne\mathcal D(k',q') \,.\label{inequality}
\end{equation}
To obtain equality, we need to dig further in the formalism of Planck-scale deformed relativistic theories.

\subsection{The momentum-dependence of Lorentz transformations}

In order to solve the problem expressed in \eqref{inequality} it may be of some help to focus on a particular deformed relativistic framework, and then try to generalize our approach. $\kappa$-Poincar\'{e} is the most-studied Hopf-algebra deformation of relativistic symmetries. This algebra fits within our general framework. $\kappa$-Poincar\'{e} possesses a so-called 'bicrossproduct structure' \cite{kMinkRelation}, meaning that both the algebra and the coalgebra are a semidirect product of a momentum sector with the Lorentz algebra, and this structure implies the existence of a co-action (or 'backreaction') of the momentum sector on the Lorentz part, which is a novelty of the model. The discovery of the bicrossproduct structure was instrumental in identifying the noncommutative spacetime this Hopf algebra acts covariantly on, the so-called $\kappa$-Minkowski spacetime \cite{kMinkRelation}. \\
Later the phenomenon of `backreaction' was given a physical interpretation \cite{GubitosiMercati}: it is the fact that, to boost in a covariant way a set of particles participating in a vertex, one needs to transform each particle momentum with a different rapidity.
This feature can be formalized by making explicit the Lorentz group element implicit in \eqref{inequality}. Call $p' = \Lambda (\Xi , p)$, where $\Xi \in SO(3,1)$, $p \in \mathcal T(p)$ and $\Lambda : SO(3,1) \times \mathcal T(p) \to \mathcal T(p)$ represents the transformation law of momentum $p$ by the Lorentz group element $\Xi$. Then we can write
\begin{equation}\label{Backreaction2}
\Lambda(\Xi, k \oplus q) =  \Lambda(\Xi^{(1)}(k, q), k) \oplus \Lambda(\Xi^{(2)}(k ,q),  q).
\end{equation}
where $\Xi_1 = \Xi_1(k,q)$ and $\Xi_2 =  \Xi_2(k,q)$ are $SO(3,1)$-valued functions of $k$, $q$ and $\Xi$ (in the special-relativistic case they both coincide with $\Xi$). In other words, as observed first in~\cite{GubitosiMercati}, while Special Relativity is implicitly assuming that the three momenta  in the vertex $k$, $q$ and $k \oplus q$ all transform with the same Lorentz group element $\Xi$, the nonlinearity of the momentum composition law may impose more complicated relations, in which each momentum transforms with a different Lorentz group element. In \cite{GubitosiMercati} the particular case of $\kappa$-Poincar\'e was studied, and the findings of~\cite{GubitosiMercati}  were that there is a `backreaction' effect of momenta on Lorentz transformations, which, for infinitesimal transformations, can be interpreted as an action of momenta on rapidity.
 The complications due to the 'backreaction' factor are the explicit manifestation of quantum deformations of the symmetry group: even if we map $(x,p)\rightarrow (X,P)$, in order to simplify the algebraic sector to the classical Poincar\'{e} algebra, we still have to deal with a complex coalgebraic sector which cannot be mapped away. This sector of the Hopf algebra describes the deformations. Therefore if on one hand one can express Lorentz transformation through classical hyperbolic functions depending on rapidity \eqref{UndeformedBoost}, on the other hand the nonlinearity of the composition law for momenta, due to the nontrivial coproduct, imposes deformed transformations on the parameter $\xi$ itself.

Those deformations can be easily calculated in general: equations \eqref{Backreaction2} impose a series of constraints on the functions $\Xi^{(1)}{}_\mu{}^\nu(k,q)$, $\Xi^{(2)}{}_\mu{}^\nu(k,q)$, $\mathcal D_\mu (k,q)$ and $\Lambda_\mu( \Xi,q)$. In~\cite{FlaJosePepe} these constraints were calculated in a perturbative setting (at first order in $M^{-1}$ and assuming undeformed rotational simmetry). The constraints found were enough to completely fix $\Xi^{(1)}{}_\mu{}^\nu(k,q)$ and $\Xi^{(2)}{}_\mu{}^\nu(k,q)$, and to establish a few relationships between the parameters of $\mathcal D_\mu (k,q)$ and $\Lambda_\mu( \Xi,q)$.\footnote{Alternatively, these constraints could be interpreted as fixing completely the parameters of $\Lambda_\mu( \Xi,q)$ as functions of the parameters of $\mathcal D_\mu (k,q)$, $\Xi^{(1)}{}_\mu{}^\nu(k,q)$ and $\Xi^{(2)}{}_\mu{}^\nu(k,q)$.} 
Using a simplified notation, equation \eqref{Backreaction2} can be expressed in a more manageable way as
\begin{equation}
\Lambda(\xi,\mathcal D(k,q))=\mathcal D(\Lambda(\xi_1,k),\Lambda(\xi_2,q))\,.\label{Backreaction1}
\end{equation}
We will apply this approach further in this article, finding a general expression for backreaction on rapidity, alongside some constraints on parameters $d_{1,2,3}$, imposed by the generalised Relativity Principle here addressed.

\section{Perturbative results up to first order in $1/M$}\label{sec:Perturb}

We perform the analysis sketched above for general deformations (Eq. \eqref{gendef} above) where $M$ is of order of Planck mass up to first order in $1/M$. In this case, $[\hat x_\mu, \hat x_\nu] = i(a_\mu \hat x_\nu - a_\nu \hat x_\mu) + \mathcal O(1/M^2)$, where $a_\mu \sim \frac1Mu_\mu$, $u_\mu u^\mu \in \{-1,0,1\}$, which corresponds to time-, light- and space-like deformations respectively. The NC coordinates are given in eq. \eqref{linreal}. We will develop our analysis from the commutation relations defined in \eqref{CommRel1} and trivial momentum space coordinates ones:
\begin{equation}
[\hat x_\mu, \hat x_\nu] = i(a_\mu \hat x_\nu - a_\nu \hat x_\mu),\quad [p_\mu, p_\nu]=0, \label{CommRel2}
\end{equation}
where the relation between vectors $a^\mu$ and $u^\mu$ is
\begin{equation}
 \quad a_\mu = \frac{c_1-c_2}Mu_\mu.
\end{equation}
From \eqref{CommRel2} and definition \eqref{linreal} we can easily obtain the deformed Heisenberg relations
\begin{eqnarray}
[p_\mu, \hat x_\nu] = -i\eta_{\mu\nu}(1+\frac{c_1}{M} (u\cdot p))-\frac{i}{M}( c_2 p_\mu u_\nu + c_3 u_\mu u_\nu (u\cdot p) + c_4 u_\mu p_\nu).
\end{eqnarray}
Another important feature of our Hopf-algebra structure, at first order in $M^{-1}$, is the generalized addition of momenta (composition law for momenta) which can be calculated from eq.\eqref{comp&copr}:
\begin{equation}
(k\oplus q)_\mu =k_\mu + q_\mu + M^{-1}\big(c_1 k_\mu(u\cdot q) + c_2 (u\cdot k) q_\mu + c_3 u_\mu(u\cdot k)(u\cdot q) + c_4 u_\mu (k\cdot q)\big).
\end{equation}

Coordinates $X_\alpha$, momenta $P_\beta$ and the Lorentz generators $M_{\mu\nu}=X_\mu P_\nu - X_\nu P_\mu$, in the basis satisfying trivial commutation relations, can be parametrized, in terms of the coordinates $(x,p)$, by a family of three parameters $d_{1,2,3}\sim 1/M$:
\begin{align}
X_\mu &= x_\mu - d_1( x_\mu(u\cdot p) -  u_\mu(x\cdot p))- 2d_3 u_\mu(u\cdot x)(u\cdot p) - 2d_2 (u\cdot x) p_\mu, \\
P_\mu &= p_\mu + d_1(u\cdot p) p_\mu + d_2 u_\mu p^2 + d_3(u\cdot p)^2 u_\mu, \\
M_{\mu\nu}&=(x_\mu p_\nu - x_\nu p_\mu)-(u_\mu x_\nu - u_\nu x_\mu)[d_2p^2+d_3(u\cdot p)^2]-[d_1 (x\cdot p) + 2 d_3(u\cdot x)(u\cdot p)](u_\mu p_\nu - u_\nu p_\mu).
\end{align}
The Casimir operator expressed in terms of $d_{1,2,3}$ is 
\begin{equation}
\mathcal C = P^2=p^2+2(d_1+d_2)(u\cdot p)p^2+2d_3(u\cdot p)^3. \label{Casimir}
\end{equation}
Generally, the coefficients $c_{1,2,3,4}$ and $d_{1,2,3}$ are in principle independent. However a stricter Hopf algebra structure will allow-us to impose a few constraints between the two sets of parameters as we will see in the nex section for the case of $\kappa$-Poincar\'{e}.

\subsection{$\kappa$-Poincar\'e Hopf algebra}
In the $\kappa$-Poincar\'e case, the following relation holds~\cite{KovMel}
\begin{equation}
[P_\mu,\hat x_\nu]=-i[\eta_{\mu\nu}(1+a\cdot P)-a_\mu P_\nu] + \mathcal O(1/M^2),\label{kPoincHeis}
\end{equation}
from which we can obtain some relations between the two set of parameters:
\begin{equation}
d_1=-\frac{c_2}{M}, \quad d_2=-\frac{c_1-c_2+c_4}{2M}, \quad d_3=-\frac{c_3}{2M}.\label{ConstrParam}
\end{equation}
Therefore in this specific case the coordinates and generators in the new basis can be  expressed in terms of parameters $c_{1,2,3,4}$ only:
\begin{align}
X_\mu &=x_\mu  + M^{-1}\big(c_2[x_\mu (u\cdot p) + u_\mu (x\cdot p)] + c_3 u_\mu (u\cdot x)(u\cdot p) + (c_1-c_2+c_4)(u\cdot x) p_\mu\big), \\
P_\mu &=p_\mu  - \frac{c_2}{M}(u\cdot p)p_\mu - \frac{u_\mu}{M}\left[
\frac{c_1-c_2+c_4}2 p^2+\frac{c_3}2 (u\cdot p)^2\right], \\  
\begin{split}
M_{\mu\nu}&=(x_\mu p_\nu - x_\nu p_\mu)+(u_\mu x_\nu - u_\nu x_\mu)\left[\frac{c_1-c_2+c_4}{2M} p^2+\frac{c_3}{2M}(u\cdot p)^2\right]+\\
&+M^{-1}[c_2 (x\cdot p) + c_3(u\cdot x)(u\cdot p)](u_\mu p_\nu - u_\nu p_\mu).
\end{split}
\end{align}
Using relations \eqref{ConstrParam} also the Casimir of the algebra \eqref{Casimir} can be re-expressed as
\begin{equation}\label{CasimirkPoinc}
\mathcal C = P^2=p^2-\frac{c_1+c_2+c_4}{M}(u\cdot p)p^2-\frac{c_3}{M}(u\cdot p)^3. 
\end{equation}
We can now easily verify that, at first order in $1/M$, for $\kappa$- Poincar\'e, the commutation relations with $M_{\mu\nu}$ are just:
\begin{align}
[M_{\mu\nu},\hat x_\lambda] &=\hat x_\mu\eta_{\nu\lambda}-\hat x_\nu\eta_{\mu\lambda}+ a_\mu M_{\nu\lambda}-a_\nu M_{\mu\lambda},
\end{align}
and that the $\kappa$-Poincar\'e Hopf algebra structure is
\begin{align}
\Delta P_\mu =& \Delta_0 P_\mu + P_\mu\otimes a\cdot P + a_\mu P_\alpha \otimes P^\alpha \,,\\
\Delta M_{\mu\nu}=& \Delta_0 M_{\mu\nu} + (\delta^\alpha_\mu a_\nu - \delta^\alpha_\nu a_\mu)P^\beta\otimes M_{\alpha\beta} \,, \\
S(P_\mu) =&-P_\mu + (a\cdot P)P_\mu - a_\mu P^2 \,,\\
S(M_{\mu\nu})=&-M_{\mu\nu}-P^\alpha(a_\mu M_{\nu\alpha} - a_\nu M_{\mu\alpha}) \,,
\end{align}
which is in agreement with the exact results obtained in Ref.\cite{KovMel}.

\subsection{Parametrizing the momentum dependence of Lorentz parameters}\label{sec:backreaction}
In order to find the most general expression for rapidities $\xi_1$ and $\xi_2$ in 1+1 dimensions at first order in $1/M$ we can write
\begin{eqnarray}
\xi_1 &=&\xi\left(1+M^{-1}(A^\alpha k_\alpha+B^\beta q_\beta)\right), \\
\xi_2 &=&\xi\left(1+M^{-1}(C^\gamma k_\gamma+D^\delta q_\delta)\right).
\end{eqnarray}
Parametrising equation \eqref{Backreaction1} we obtain many interesting results. First of all we find that the rapidity parameter related to one particle is not affected by its own momentum, namely $A=D=0$.
 Moreover, we find a solution for the other parameters, obtaining the general solution for momentum-dependent rapidities that we were looking for:
\begin{equation}
\begin{split}
&\xi_1 =  \xi\left(1-(M^{-1}c_2+d_1)(u\cdot q) \right), \\
&\xi_2 =  \xi\left(1-(M^{-1}c_1+d_1)(u\cdot k) \right).
\end{split}
\end{equation}
Finally, the Relativity Principle implemented in \eqref{Backreaction2} and \eqref{Backreaction1} in addition to the aforementioned results, implies two constraints linking the momenta parameters $d_{1,2,3}$ with the coordinate parameters $c_{1,2,3,4}$:
\begin{equation}
 d_1+d_2=-\frac{c_1+c_2+c_4}{2M}, \quad d_3=-\frac{c_3}{2M}. \label{BackConstraints}
\end{equation} 
It should be noticed that these latter constraints, while on one hand leave one free parameter, on the other hand make it possible to re-express the Casimir of the algebra \eqref{Casimir} completely in terms of coordinate parameters $c_{1,2,3,4}$, since $d_1$ and $d_2$ only appear as a sum in $\mathcal{C}(p)$. Notably, the Casimir we obtain in this general case has the same form of the $\kappa$-Poincar\'{e} one \eqref{CasimirkPoinc}.

In order to specialize our general formulas to the $\kappa$-Poincar\'e backreaction we just need to fix $d_1=-c_2/M$, since two of the constraints in \eqref{ConstrParam} are already contained in \eqref{BackConstraints}:
\begin{equation}
\begin{split}
& \xi_1 =  \xi, \\
& \xi_2 =  \xi\left(1-\frac{c_1-c_2}{M}(u\cdot k)\right) = \xi(1-a\cdot k).
\end{split}
\end{equation}
These expressions reproduce at first order the exact result for backreaction factors in $\kappa$-Poincar\'e, obtained in~\cite{GubitosiMercati}.

Other combinations that may be worthy of some further investigation are the ones in which some of the parameters are zero. For instance when $d_1=d_2=d_3=0$ of course boost and Casimir are classical, but we would still have nontrivial momenta coproducts, deformed composition law for momenta and momentum-dependent rapidity, depending on two free parameters, say $c_1$ and $c_2$, while $c_4=-(c_1+c_2)$. This particular noncommutative spacetime would have classical Poincar\'{e} symmetries, but nontrivial coalgebra and a deformed Relativity Principle formalized by  
\begin{equation}
\begin{split}
&\xi_1 =  \xi\left(1-M^{-1}c_2(u\cdot q) \right), \\
&\xi_2 =  \xi\left(1-M^{-1}c_1(u\cdot k) \right). 
\end{split}
\end{equation}

The case $c_1=c_2=c_3=c_4=0$ would still have an undeformed Poincar\'{e} algebra Casimir, trivial composition law for momenta and commutative coordinates, but also a deformed boost generator depending on just one free parameter $d_1=-d_2$:
\begin{equation}
M_{\mu\nu}=(x_\mu p_\nu-x_\nu p_\mu)+d_1\left[(u_\mu x_\nu -u_\nu x_\mu)p^2-(x\cdot p)(u_\mu p_\nu -u_\nu p_\mu)\right],
\end{equation}
For this reason this commutative model would still need  momentum-dependence of Lorentz parameters in order to conserve particle interaction vertices. The rapidity parameters in this case assume a nice symmetric form:
\begin{equation}
\begin{split}
&\xi_1 =  \xi\left(1-d_1(u\cdot q) \right), \\
&\xi_2 =  \xi\left(1-d_1(u\cdot k) \right). 
\end{split}
\end{equation}

In the case where coordinates $\hat x_\mu$ are commutative (at least in the first order), which is given by $c_1=c_2\equiv c$, the Casimir and the boost operator are deformed, depending on $c$, $c_3$ and $c_4$
\begin{align}
\mathcal C &= P^2=p^2-\frac{2c+c_4}{M}(u\cdot p)p^2-\frac{c_3}{2M}(u\cdot p)^3, \\
M_{\mu\nu}&=(x_\mu p_\nu - x_\nu p_\mu)+(u_\mu x_\nu - u_\nu x_\mu)\left[\frac{c_4}{2M} p^2+\frac{c_3}{2M}(u\cdot p)^2\right] + M^{-1}[c (x\cdot p) + c_3(u\cdot x)(u\cdot p)](u_\mu p_\nu - u_\nu p_\mu),
\end{align}
but the rapidity parameters are also symmetric like in the previous case
\begin{equation}
\begin{split}
&\xi_1 =  \xi\left(1-(M^{-1}c+d_1)(u\cdot q) \right), \\
&\xi_2 =  \xi\left(1-(M^{-1}c+d_1)(u\cdot k) \right). 
\end{split}
\end{equation}

The study of $M$-deformed particle interaction vertices is of paramount importance to characterize the relativistic framework determined by the Hopf-algebra formalism related to nontrivial coordinate commutation relations.  Momentum-dependence of Lorentz transformations, as we explained in section~\ref{sec:GLT}, is an essential feature in this picture, whose role is to match Hopf-algebra coproducts and composition laws \eqref{comp&copr} with a set of reference frames transformations, characterized not just by the observers spacetime positions (and their relative velocity), but also by the energy of the particles that they observe. This picture has inspired a rich phenomenology, investigating a class of apparent nonlocal effects arising from the comparison of signals measured by distant observers.

\section{Phenomenological application to the study of astrophysical signals}\label{sec:Phenom}

There is a large literature studying the possibility to observe time delays between particles with different energies emitted by astrophysical sources \cite{phenomenology,neutrini1}. Such an effect is very well known to be related to timelike (e.g. $\kappa$-Minkowski) coordinate noncommutativity  \cite{kMinkRelation}. There are also some preliminary results investigating Planckian {\it transverse effects} (effects on the direction of propagation, rather than time-of-arrival) also called {\it dual gravity lensing} \cite{transverse,neutrini2} which in some cases seem to be related to noncommutative spacetime models already studied for their non-Pauli statistics \cite{nonPauli}.

\begin{figure}
 \centering
\fbox{\includegraphics[scale=0.28]{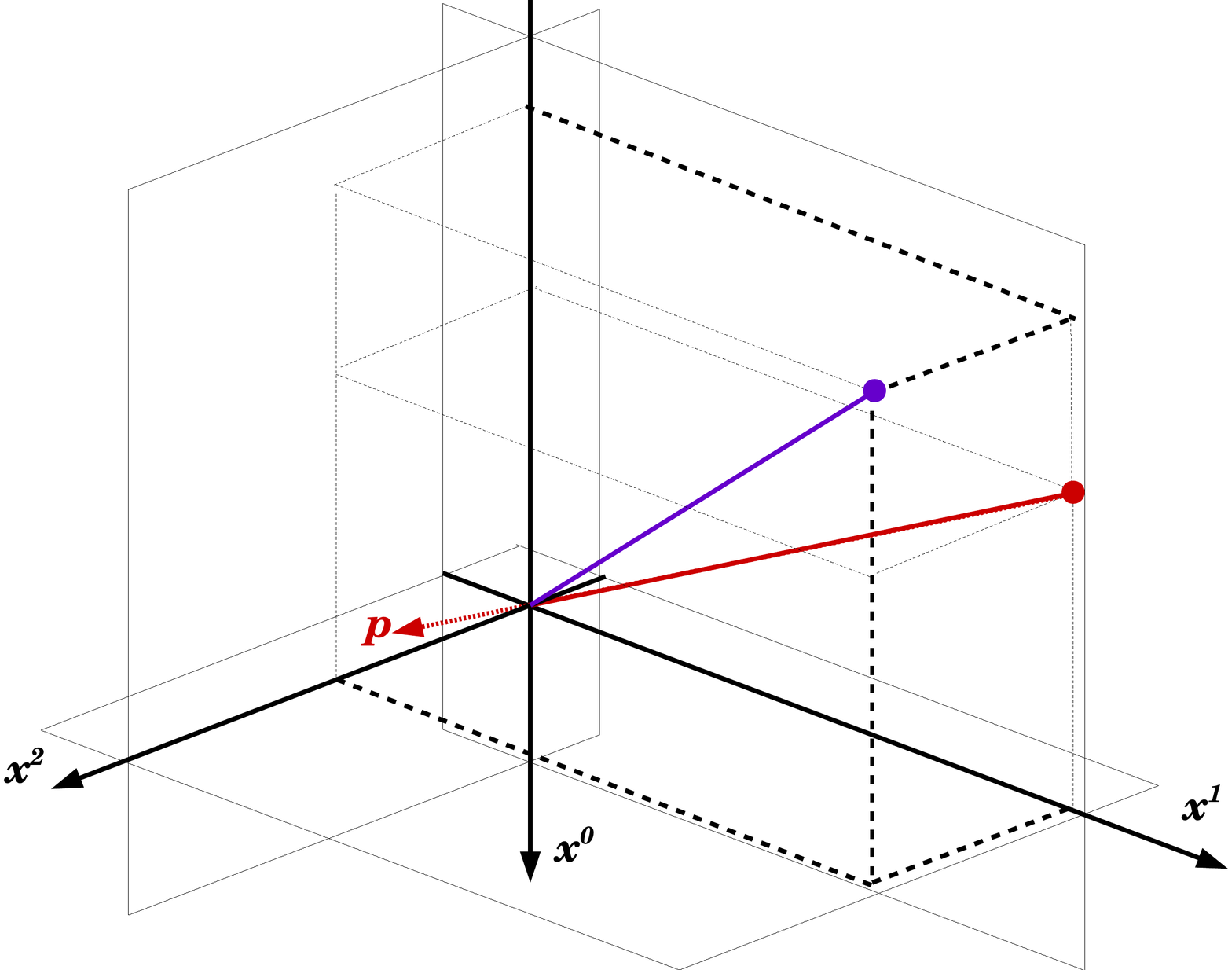}}
\caption{\footnotesize This simple scheme shows the characteristics of the effects that an observer would measure in the most general case. Observing some particle with momentum $p=|p_1|$, such an observer would infer, according to undeformed (classical) mechanics, the particle's trajectory to follow the red line, while taking into account the deformation effects, the emission time and the direction of propagation (and therefore the position of the source) may change according to the particle's energy, as described by the violet line.}
\label{fig:schema}
\end{figure}

An interesting aspect of the vectorial generalization presented here is the possibility to summarize the entire range of spacetime-quantization phenomenological features on particle propagation. We will formalize all those different effects within the {\it Relative Locality} framework \cite{relloc,kbob}, in order to  intuitively identify the physical interpretation of the different kind of Planck-scale deformations. This is possible since Relative Locality is a classical framework in which both quantum and spacetime curvature  effects are turned off, i.e. $\hslash\rightarrow 0$ and $G\rightarrow 0$. At the same time however Planck-scale   deformations are still present as modification to the classical Hamiltonian mechanics since in this limit $M\sim M_P=\sqrt{G/\hbar}\neq 0$. Since we are now working within a classical framework there is no point in describing spacetime using noncommuttive coordinates $\hat{x}^\alpha$. We can just use commutative ones $x^\alpha$, while all the features so far formalized using commutation relations will be, from now on, expressed as nontrivial Poisson brackets, following the well known relation 
\begin{equation}
\left[A,B\right]=i\hslash \{A,B\}.
\end{equation}
The other fundamental object for our analysis is the Hamiltonian operator that we will assume to be the Casimir (\ref{Casimir}) of the algebra, (see for more details Ref. \cite{kbob,altriesempi})
\begin{equation}
{\cal C}=p^2-\frac{c_1+c_2+c_4}{M}(u\cdotp p)p^2-\frac{c_3}{M}(u\cdotp p)^3,
\end{equation}
in which parameters $d_2$ and $d_3$ of \eqref{Casimir} have been re-expressed using constraints \eqref{BackConstraints} obtained in Section \ref{sec:backreaction}.
We can formalize the deformation effects to the particles' propagation through spacetime from $\bar x$ to $x$, by studying their worldlines
\begin{equation}
\int_{\bar{x}^i}^{x^i}dx^i=\int_{\bar{x}^0}^{x^0} v^i dx^0, \label{worldlines}
\end{equation}
where, as in classical Hamiltonian mechanics, $v^i\equiv \dot{x}^i/\dot{x}^0$ is the particle velocity along direction $i$, and 
\begin{equation}
\dot{x}^\alpha=\{\mathcal{C},x^\alpha\}. \label{quadrivel}
\end{equation}
 Let us assume now to observe a particle arriving at the origin with momentum $|p|\equiv p_1$ in 2+1 dimensions, i.e. we set the ``1" direction as the arriving direction of the particle and $p_2=0$ (see Figure \ref{fig:schema}). In classical mechanics equation (\ref{quadrivel}) would give us $\dot{x}^2=0$, and then the velocity we have in the worldlines (\ref{worldlines}) would only have one component $v^1\neq 0$. We may then trace back the source of the particle along direction ``1" at a distance $\Delta x^1=v^1\cdotp \Delta x^0$. However since in our case the relation between momenta and velocity is in general not linear anymore, what can we say about the spacetime position of the source and travel time? 

In order to quantify the different effects we get in the timelike, spacelike and lightlike cases, let us take into account a massless particle whose on-shell relation ${\cal C}=0$ gives
\begin{equation}
|p_1|\simeq p_0\left(1-\frac{c_3}{2 M}p_0(u^0-u^1)^3\right).
\end{equation}
After a few simple calculations we can find that in the general case the following relations hold
\begin{equation}
\left\{\begin{array}{l}
v^1=1+\frac{c_3}{M}p_0(u^0-u^1)^3\\
v^2=-\frac{3}{2}\frac{c_3}{M}p_0 u^2(u^0-u^1)^2
\end{array}\right. .
\end{equation}

As we previously mentioned, in the most studied example in literature in which we have just timelike deformation $u^0=1$, $u^i=0$, the only physical observable effect would be a time delay between particles with different energies, given the deformation of $v^1$. High energy particles would then propagate faster (or slower, according to the sign of $c_3$) than the low energetic ones. On the other hand the particle direction of propagation can always be inferred by an observer, since we have $u^2=0$ and then no $v^2$ component. However in the other two cases (lightlike and spacelike) the situation is a little bit more complicated: simultaneously-emitted particles with different energies would not only arrive at different times at the detector, but the observer would infer a very small apparent angle between the two directions of propagation of the order of $\theta\simeq v^2\sim p_0/M_P$.

\section{Discussion and outlook}

In this paper we have presented a study of a large set of noncommutative spacetimes, their symmetries and their physical features. We developed our investigation starting from the parametrization presented in \eqref{linreal}, evolving the proposal suggested in \cite{EPJC}. We have studied the Hopf algebra structure related to the symmetry generators of a vector-like generalization of $\kappa$-Minkowski spacetime, as well as their relativistic framework. We have found that the relations between the parameters obtained from particles vertices conservation-laws \eqref{BackConstraints} are compatible with the ones independently found imposing the characteristic $\kappa$-Poincar\'{e} Heisenberg-algebra commutation relations \eqref{kPoincHeis}. The $\kappa$-Poincar\'{e} Hopf algebra is in fact a specific subset of this general family of relativistic algebras, characterized by one more constraint between parameters $c_{(1,2,3,4)}$ and $d_{(1,2,3)}$.

We have discussed the momentum-dependent Lorentz transformations that one should in general expect  when studying particles interactions, obtaining a generic expression for momentum-dependence of the boost parameters of different particles. We also analysed a few interesting particular cases in which even if the symmetry generators algebra is trivial or coordinates commute, non trivial effects and momentum-dependence on the boost rapidity should be nonetheless expected due to the nontrivial composition law for momenta.
 More insight on the theoretical aspects of this effect could come from the study of classical quasigroups with generalized composition law of parameters (see e.g. \cite{Batalin}) which may be useful for the task of composing nonlinear realizations as considered in this paper.
 
Once we defined the relativistic kinematics, we have finally been able to investigate on the phenomenology of particle propagation along astrophysical distances, connecting the shape of different noncommutative spacetimes with a characteristic phenomenology. Both the coordinates nontrivial commutation relation and the Planck-scale deformed particles velocities are in fact formalized through their typical vector $u^\mu$ configuration (timelike, lightlike or spacelike). The vector-like generalization in fact allows us to summarize with a simple parametrization a large range of effects (see \cite{phenomenology} and \cite{transverse}) previously only studied separately.

The phenomenological interpretation is particularly manageable in the so-called Relative Locality limit (in which both $G,\hslash\rightarrow 0$), where phase space is deformed by Planck-scale Hopf algebraic remnants. It is remarkable to notice that this classical limit finds a formal realization (at least at first order in $1/M$) in the so called {\it Finsler geometry} framework \cite{Finsler}, a generalization of Riemannian geometries whose fumdamental element is a non-quadratic norm $F(\dot{x})$ instead of a quadratic arclength. Then it may be interesting to dedicate some further effort in studying a large family of those geometries, their metrics, and their connections, characterized by different shapes of vector $u^\mu$.

\section*{Acknowledgements}
The work by S.M. and D.P. has been fully supported by Croatian Science Foundation under the Project No. IP-2014-09-9582. N.L. acknowledges support by the European Union Seventh Framework Programme (FP7 2007-2013) under grant agreement 291823 Marie Curie FP7-PEOPLE-2011-COFUND (The new International Fellowship Mobility Programme for Experienced Researchers in Croatia - NEWFELPRO), and also partial support by the H2020 Twinning project n$^\text{o}$ 692194, RBI-TWINNING and by the 000008 15 RS {\it Avvio alla ricerca} 2015 fellowship (by the Italian ministry of university and research). F.M. was supported by Perimeter Institute for Theoretical Physics. Research at Perimeter Institute is supported by the Government of Canada through Industry Canada and by the Province of Ontario through the Ministry of Research and Innovation.

\end{document}